\documentstyle[12pt]{article}

\begin{document}
% **************************** MACROS

\newcommand{\gi}{g_{i}}
\newcommand{\gj}{g_{j}}

\newcommand{\gim}{g_{i+1}}
\newcommand{\gjm}{g_{j+1}}

\newcommand{\gime}{g_{i-1}}
\newcommand{\gjme}{g_{j-1}}

\newcommand{\ai}{a_{i}}
\newcommand{\aj}{a_{j}}

\newcommand{\aim}{a_{i+1}}
\newcommand{\ajm}{a_{j+1}}

\newcommand{\bi}{b_{i}}
\newcommand{\bj}{b_{j}}

\newcommand{\bmasi}{b_{i}^{+}}
\newcommand{\bmasj}{b_{j}^{+}}

\newcommand{\bmasim}{b_{i+1}^{+}}
\newcommand{\bmasjm}{b_{j+1}^{+}}

\newcommand{\bmei}{b_{i}^{-}}
\newcommand{\bmej}{b_{j}^{-}}

\newcommand{\bmeim}{b_{i+1}^{-}}
\newcommand{\bmejm}{b_{j+1}^{-}}

\newcommand{\bzei}{b_{i}^{0}}
\newcommand{\bzej}{b_{j}^{0}}

\newcommand{\bzeim}{b_{i+1}^{0}}
\newcommand{\bzejm}{b_{j+1}^{0}}

\newcommand{\ball}{b_{i}^{+,-,0}}

\newcommand{\gii}{g_{i}^{-1}}
\newcommand{\gjj}{g_{j}^{-1}}

\newcommand{\giim}{g_{i+1}^{-1}}
\newcommand{\gjjm}{g_{j+1}^{-1}}

\newcommand{\giime}{g_{i-1}^{-1}}
\newcommand{\gjjme}{g_{j-1}^{-1}}

\newcommand{\Gi}{G_{i}}
\newcommand{\Gj}{G_{j}}

\newcommand{\Gim}{G_{i+1}}
\newcommand{\Gjm}{G_{j+1}}

\newcommand{\Gime}{G_{i-1}}
\newcommand{\Gjme}{G_{j-1}}

\newcommand{\Ai}{A_{i}}
\newcommand{\Aj}{A_{j}}

\newcommand{\Aim}{A_{i+1}}
\newcommand{\Ajm}{A_{j+1}}

\newcommand{\Bi}{B_{i}}
\newcommand{\Bj}{B_{j}}

\newcommand{\Bmasi}{B_{i}^{+}}
\newcommand{\Bmasj}{B_{j}^{+}}

\newcommand{\Bmasim}{B_{i+1}^{+}}
\newcommand{\Bmasjm}{B_{j+1}^{+}}

\newcommand{\Bmei}{B_{i}^{-}}
\newcommand{\Bmej}{B_{j}^{-}}

\newcommand{\Bmeim}{B_{i+1}^{-}}
\newcommand{\Bmejm}{B_{j+1}^{-}}

\newcommand{\Bzei}{B_{i}^{0}}
\newcommand{\Bzej}{B_{j}^{0}}

\newcommand{\Bzeim}{B_{i+1}^{0}}
\newcommand{\Bzejm}{B_{j+1}^{0}}

\newcommand{\Ball}{B_{i}^{+,-,0}}

\newcommand{\Gii}{G_{i}^{-1}}
\newcommand{\Gjj}{G_{j}^{-1}}

\newcommand{\Giim}{G_{i+1}^{-1}}
\newcommand{\Gjjm}{G_{j+1}^{-1}}

\newcommand{\Giime}{G_{i-1}^{-1}}
\newcommand{\Gjjme}{G_{j-1}^{-1}}

\newcommand{\lmasu}{L^{\left( 1 \right) }_{+}}
\newcommand{\lmasd}{L^{\left( 2 \right) }_{+}}
\newcommand{\lmeu}{L^{\left( 1 \right) }_{-}}
\newcommand{\lmed}{L^{\left( 2 \right) }_{-}}

\newcommand{\lzd}{L^{\left( 2 \right) }_{0}}
\newcommand{\lzu}{L^{\left( 1 \right) }_{0}}

\newcommand{\ltimesu}{L_{\times}^{(1)}}
\newcommand{\ltimesd}{L_{\times}^{(2)}}
\newcommand{\ltimes}{L_\times}
\newcommand{\lmast}{L_{\times\!\!\!|}^{+}  }
\newcommand{\lmet}{L_{\times\!\!\!|}^{-}  }
\newcommand{\lzt}{L_{\times\!\!\!|}^{0}  }
\newcommand{\lpmt}{L_{\times\!\!\!|}^{\pm}  }

\newcommand{\da}{ {\cal D}A }
\newcommand{\wilson}{\langle W(\gamma) \rangle}
\newcommand{\wilsonpm}{\langle W( L_{\pm}) \rangle}
\newcommand{\wilsonz}{\langle W( L_{0}) \rangle}
\newcommand{\wilsont}{\langle W( L_{\times}) \rangle}
\newcommand{\wilsonmas}{\langle W( L_{+}) \rangle}
\newcommand{\wilsonmenos}{\langle W( L_{-}) \rangle}

\newcommand{\wilsonzh}{\langle W(\hat L_{0}) \rangle}
\newcommand{\wilsonth}{\langle W(\hat L_{\times}) \rangle}
\newcommand{\wilsonmash}{\langle W(\hat L_{+}) \rangle}
\newcommand{\wilsonmenosh}{\langle W(\hat L_{-}) \rangle}
\newcommand{\wilsonpmh}{\langle W( L_{\pm}) \rangle}

\newcommand{\wilsonr}{\langle W( R_{\pm}) \rangle}
\newcommand{\wilsonb}{\langle W(\lmast) \rangle}

\newcommand{\wilsonbpm}{\langle W(\lpmt) \rangle}

\newcommand{\wilsong}{\langle W(\lmasd \lmeu  ) \rangle}

\newcommand{\wilsongg}{\langle W( \lmasd \lmasd  ) \rangle}
\newcommand{\wilsonx}{\langle W(\lmasu \lmasd  ) \rangle}
\newcommand{\wilsony}{\langle W( \lmasd \lmasu ) \rangle}

\newcommand{\wilsongfg}{\langle W(\lmasd \lmasu \lmasd ) \rangle}

\newcommand{\wilsoni}{\langle W(\lzd  ) \rangle}
\newcommand{\wilsongi}{\langle W(\lmasd ) \rangle}
\newcommand{\wilsongii}{\langle W(\lmed  ) \rangle}
\newcommand{\wilsona}{\langle W(\ltimesd ) \rangle}

\newcommand{\wdume}{\langle W(\lmasd \lmeu  ) \rangle}

\newcommand{\chern}{e^{iS_{CS}}}
\newcommand{\ep}{\epsilon_{\mu\nu\rho}}
\newcommand{\epp}{\epsilon_{\mu\nu\rho}}
\newcommand{\dx}{dx^{\mu}}
\newcommand{\dy}{dy^{\nu}}
\newcommand{\dz}{dz^{\rho}}
\newcommand{\deriv}{\frac{\delta}{\delta A_{\rho}^{a}(x)}}
\newcommand{\integral}{Z^{-1} \int}
\newcommand{\cuatro}{\frac{-i4\pi}{k}}
\newcommand{\dos}{\frac{-i2\pi}{k}}
\newcommand{\tra}{{\cal T}r}

% ************************************* 0
\begin{titlepage}
%\rightline{Preprint 92\ \ \ \ \ \ \ \ \ \ }
\vspace{0.5cm}
\begin{center}
          {\LARGE    Link invariants for intersecting \\[0.25cm]
                     loops}

\end{center}

\vspace{0.5cm}
\begin{center}
     {\small\bf   Daniel Armand Ugon, Rodolfo Gambini,
                  and Pablo Mora}
\end{center}
\vspace{0.5cm}

\begin{center}
         {\small Instituto de F\'{\i}sica, Facultad de Ciencias,
                 Trist\'an Narvaja 1674, and \\
                 Instituto de F\'\i sica, Facultad de Ingenier\'\i a,
                 J. Herrera y Reissig 565, C. C. 30 \\
                 Montevideo, Uruguay.}
\end{center}

\vspace{0.5cm}
\begin{center}October 1992\end{center}
\vspace{0.5cm}

\begin{abstract}

We generalize  the  braid  algebra  to  the  case  of  loops  with
intersections. We  introduce  the  Reidemeister  moves  for  4  and
6-valent vertices to  have  a  theory  of  rigid  vertex
equivalence. By considering representations of the  extended  braid
algebra, we  derive  skein   relations   for   link
polynomials, which allow us to generalize any link Polynomial  to  the
intersecting case. We
perturbatively show that the HOMFLY Polynomials  for  intersecting
links  correspond
to the vacuum expectation value of the Wilson line operator of the
Chern Simon's Theory.

We make contact with quantum gravity by showing
that these polynomials are simply related with some  solutions  of
the complete set  of  constraints   with   cosmological   constant
$\Lambda $,for
loops including triple  self intersections. Previous  derivations  of
this result were restricted to the 4-valent case.
\end{abstract}
\end{titlepage}
\newpage

% ************************************* 1
%\section{\bf Introduction}

The discovery  of  a  new  set  of  canonical  variables  for  3+1
dimensional general relativity
\cite{Aste} has led to the  introduction  of  a  loop  representation
\cite{SmoRo} for quantum gravity
in  which  wave  functions  are  functionals  of  loops.  Such   a
representation was known to
exist  for  several  other  gauge  theories  such   as   Maxwell's
electromagnetism \cite{ref2} and
Yang Mills theories in the continuum \cite{ref3} and the lattice \cite{ref4}.
Jacobson and
Smolin \cite{ref5} noticed that diffeomorphism  invariant  classes  of
non intersecting loops
satisfied  all  the  constrained  equations  of  quantum  gravity.
But, it was soon
recognized that these solutions led to degenerate metrics  and  if
one considers the
determinant of the metric acting on these  wave  functions  it  is
identically zero
everywhere. It is very easy to check \cite{gamPu}  that  if  a  quantum
state is annihilated by the
determinant of the metric and the hamiltonian of the vacuum theory,
 it is also a
quantum state of an arbitrary value of the cosmological  constant.
Since one expects
different behaviors depending on the  value  of  the  cosmological
constant,  it is quite
clear that the non intersecting loop dependent states are  just  a
small and degenerate
sector  of  the  theory.  Furthermore,  the   vanishing   of   the
determinant of the metric
imposes important restrictions to the coupling of matter with  the
gravitational field
\cite{ref7} and leads to  manifolds with vanishing volume.

Non degenerate  solutions  involve  at  least  three  intersecting
loops. The first non degenerate
solution with triple self intersections was obtained some time ago
\cite{gamPu} in
terms of the
second coefficient of the Alexander-Conway  polynomial.  After
that it was shown
that a knot polynomial  closely  related  to  the  Jones
polynomial belongs to the
physical state space of quantum gravity with cosmological constant
\cite{Pullin}.
For
each  order  in  the  cosmological constant there  appear  portions  of  the
coefficients that are
annihilated by the vacuum hamiltonian constrained and not  by  the
determinant of the
metric. This suggests  a  deep  and  beautiful  connection  between
quantum gravity and
knot theory at the dynamical level.  Thus,  there  are  compelling
reasons to develop
mathematical techniques for  dealing  with  links  admitting  self
intersections.

Some
results for 4-valent intersections have been recently developed by
Kauffman \cite{Kauff},
Vassilev \cite{Vassilev}, Bar-Natan \cite{Natan}, Birman and  Lin
\cite{Lin} and Baez \cite{Baez}. In this paper we
generalized  the  braid  algebra  in  order   to   include   rigid
intersections.  We
consider the  case  of  4-valent  and  6-valent  vertices  and  we
introduce the corresponding
Reidemeister  moves  to  have  a  theory   of   vertex
equivalence. Then, we consider
a  representation  of   the   extended   braid   algebra.   This
representation leads to a set of
simple  Skein  relations  that  allow  to  generalized  any   link
polynomial to the intersecting
case. Finally we show perturbatively that the Wilson loop  average
of the Chern
Simons
Theory \cite{witten} \cite{guad} \cite{smolin} is connected with the HOMFLY
polynomial, even  for
multiple intersections. This computation also  allows  to  confirm
that the Jones
Polynomial solves the quantum gravity constraint with cosmological
constant,
for
loops including 4 and 6  valent  vertices.  Previous  proofs \cite{gamPu}
were
restricted to the 4 valent
case. We conclude with  some  remarks  concerning  other  possible
applications of the
link polynomials for intersecting loops in Quantum Gravity.

% ************************************* 2
% \section{\bf  Intersecting links invariants}

First we  will  introduce  an  extension  of  the  braid
algebra in order to include 4-valent and 6-valent intersections. The
corresponding Reidemeister
moves will be
expressed in algebraic form. However we shall  see  that  it  is  straight
forward to generalize any
oriented non  intersecting  link  invariant  to  the  intersecting
case, by using only the planar graph representation of the generalized
Reidemeister moves and without making reference to the braids. The $N$ strands
braid
group $B_{N}$  is generated by N-1 elements $g_{i} \ i=1 \ldots N-1$
  that  satisfy  the  braid
group
algebra

%EQUACION 1

\begin{eqnarray}
\gi \gii &=& \gii \gi = I \nonumber   \\
\gi \gim \gi &=& \gim \gi \gim \label{eq1}  \\
\gi \gj &=& \gj \gi  \mbox{ if $ \left| i-j \right| > 1$} \nonumber
\end{eqnarray}

This algebra reflects the invariance under Reidemeister moves of  type  II
and III. The $\gi$ or $ \gii $ generate the crossing of the ith
strand over (under ) the (i+1)th.
The braid algebra can
be extended to include rigid 4-valent interesections  of  strands.
This is done by
including the N-1 elements $\ai ,\ i=1 \ldots N-1$ ( see fig. 1).

The corresponding Reidemeister moves for the new  generators  have
the following algebraic form

%EQUACION 2

 \begin{eqnarray}
 \ai \gi &=& \gi \ai  \nonumber  \\
 \gii \aim \gi &=& \gim \ai \giim \nonumber \\
 \gi \aj &=& \aj \gi \mbox{ if $ \left| i-j \right| > 1 $} \label{eq4} \\
 \ai \aj &=& \aj \ai \mbox{ if $ \left| i-j \right| > 1 $}  \nonumber
\end{eqnarray}

For the case of 6-valent intersections the crucial point is to note
that there are three
diffeomorphic invariant ways to glue three  strands  at  one  point.
They correspond to the
three possible values of the sign of the volume  element  expanded
by the tangent
vectors at the strands in the intersecting point, i.e.  +1,-1  or
0. There are then 3(N-2)
new elements that must be included in the extended braid  algebra.
We called them $\bmasi \ , \bmei ,\bzei $ with $i=2 \ldots N-1$  (see fig. 1)

These new generators satisfy the following relations, which  follow
from the generalized
Reidemeister moves for this case.

%ECUACION 3
 \begin{eqnarray}
 \bmei &=& g_{i-1}\bmasi\gii  \nonumber \\
 \bmei &=& \gii \bmasi g_{i-1}  \nonumber \\
 \ball &=& g_{i-1} \gi \gim \ball \giim \gii g_{i-1}^{-1} \nonumber \\
 \bi \gj &=& \gj \bi \mbox{  if $j > i+1$ or $j < i-2$}  \label{eq5} \\
 \bi \aj &=& \aj \bi \mbox{  if $j > i+1$ or $j < i-2$} \nonumber \\
 \bi \bj &=& \bj \bi \mbox{ if $ \left| i-j \right| > 2 $} \nonumber \\
 \bzei &=& \gi \gime \gi \bzei \gii \giime \gii \nonumber
\end{eqnarray}

When intersections are included the algebra generated by $\gi$, $\ai$  and
$b_i$ is no longer a group.
It is now straightforward to prove that for each representation of
the $g_i'$s given by
matrices $\Gi's$ satisfying eq. \ref{eq1} , the following matrices satisfy
the extended algebra.

 \begin{eqnarray}
A_i &=&  \alpha_1 G_i + \alpha_2 G_i^{-1} + \alpha_3 I_i  \label{eq2}
\nonumber \\
B_i^{+} &=&  \beta_1 \Gi\Gime\Gi +\beta_2 \left\{ \Gi\Gime   + \Gime\Gi +
\Gi^{2} \right\}
+ \beta_3 \Gi + \beta_4 \Giime \\
B_i^{-} &=& \beta_1 \Gi\Gime\Gi + \beta_2 \left\{ \Gi\Gime + \Gime\Gi +
\Gime^{2}
\right\} + \beta_3 \Gime + \beta_4 \Gii \nonumber \\
\Bzei &=& \beta_5 \left\{ \Bmasi + \Bmei \right\} + \beta_6 I_{i-1} I_{i}
\nonumber
\end{eqnarray}

Where the coefficients $\alpha_i$   and $\beta_i $  are arbitrary complex
parameters.
If the fundamental
matrices $\Gi$  satisfy  the  usual  identities  leading  to   Skein
relations

%ECUACION 5
\begin{equation}
c_1 \Gi + c_2 \Gii + c_3 I_i = 0
\end{equation}

it is possible to
express one of the terms in eq.\ref{eq2} as a linear  combination  of  the
others.

 Any invariant polynomial $F(L)$  for oriented non intersecting  links  may
be generalized to the
intersecting case by adding the following skein relations:

%ECUACION 6
% \begin{eqnarray}
%F(A_2) &=& \alpha_1 F(G_i) + \alpha_2 F(G_i^{-1}) + \alpha_3 F(I_i) 
%\nonumber \\
%F(\ltimes) &=& \alpha_1 F(L_{+}) + \alpha_2 F(L_{-}) + \alpha_3 F(L_{0})
%\nonumber \\
%F(B_i^{+}) &=& \beta_1 F(\Gi\Gime\Gi) + \beta_2 \{ F(\Gi\Gime) + \nonumber \\
%           & & F(\Gime\Gi) + F(\Gi^{2}) \} + \beta_3 F(\Gi) + \beta_4
% F(\Giime)
%            \nonumber \\
%F(B_i^{-}) &=& \beta_1 F(\Gi\Gime\Gi) + \beta_2 \{ F(\Gi\Gime) \label{eq3}
%         \label{next} \\
%           & & + F(\Gime\Gi) + F(\Gime^{2}) \} + \beta_3 F(\Gime) + \beta_4
% F(\Gii)
%           \nonumber \\
%F(\Bzei) &=& \beta_5 \left\{ F(\Bmasi) + F(\Bmei) \right\} + \beta_6
%F(I_{i-1} I_{i})
%\nonumber
%\end{eqnarray}

 \begin{eqnarray}
F(\ltimes) &=& \alpha_1 F(L_{+}) + \alpha_2 F(L_{-}) + \alpha_3 F(L_{0})
\nonumber \\
F(\lmast) &=& \beta_1 F(\lmasd \lmasu \lmasd) + \beta_2 \{ F(\lmasd \lmasu) +
\nonumber \\
           & & F(\lmasu \lmasd) + F(\lmasd \lmasd) \} + \beta_3 F(\lmasd) +
\beta_4 F(\lmeu)
            \nonumber \\
F(\lmet) &=& \beta_1 F(\lmasd \lmasu \lmasd ) + \beta_2 \{ F(\lmasu \lmasd   )
\label{eq3}
         \label{next} \\
           & & + F(\lmasd \lmasu ) + F(\lmasu \lmasu) \} + \beta_3 F(\lmasu) +
\beta_4 F(\lmed)
           \nonumber \\
F(\lzt) &=& \beta_5 \left\{ F(\lmast) + F(\lmet) \right\} +
\beta_6 F(\lzu \lzd  )
\nonumber
\end{eqnarray}

The notation used is explained in fig. 2.
To each regular  (ambient  )  invariant   polynomial   in   the   non
intersecting case there will correspond a
regular (ambient) invariant polynomial in the  generalized  case.
By hypothesis the link invariant $F(L)$ does not change under the usual
Reidemeister moves. Hence we can prove the invariance under the generalized
moves assuming the relations (6) only by using the usual Reidemeister moves in
the planar graph form. This proof requires a sequence of transformations on
planar diagrams entirely analogous to the algebraic calculation necessary to
check that the operators in (4) satisfy the algebra (3). Thus, our
generalization
of $F(L)$ does not depend on how was $F(L)$ constructed.
We are going to apply this construction to the case of the regular
invariant polynomial $S_{L}(a,b,z)$  defined by the Skein relations:

%ECUACION 7
 \begin{eqnarray}
 S(\hat{L}_{\pm}) &=& a^{\pm 1}S(\hat{L}_{0}) \nonumber \\
 bS(L_{+}) - b^{-1}S(L_{-}) &=& zS(L_{0}) \\
 S(U) &=& 1 \mbox{\  where $U$ is the unknotted.} \nonumber
 \end{eqnarray}

The $S_{L}(a,b,z) $ is related with the ambient HOMFLY polynomial
$P_{L}(a,b,z)$  by
\begin{equation}
P_{L}(t=ab,z)= a^{-w(L)}S_{L}(a,b,z)
\end{equation}

where $w(L)$ is the writhe of the  link  $L$.  The  HOMFLY  polynomial
contains as a particular case for $z= t^{1/2}-t^{-1/2} $ the  Jones
Polynomial.  It  is
  now  trivial   to
generalize the $S_{L}$ polynomial to the intersecting case by adding
the variables $\alpha_i \ i=1 \ldots 3$ and $\beta_i \ i=1,\ldots 6$.
 From eq.\ref{eq4} and eq.\ref{next}   we
obtain  the  following  regular  invariant
polynomial $S_{L}(a,b,z,\alpha_i,\beta_j)$ defined by

%Ecuacion 8
\begin{center}
 \begin{eqnarray}
 S(\hat{L}_{\pm}) &=& a^{\pm}S(\hat{L}_{0}) \nonumber \\
 bS(L_{+}) - b^{-1}S(L_{-}) &=& zS(L_{0}) \nonumber \\
 S(U) &=& 1 \mbox{  where $U$ is  unknotted.} \nonumber \\
S(\ltimes) &=& \alpha_1 S(L_{+}) + \alpha_2 S(L_{-}) + \alpha_3 S(L_{0})
\nonumber \\
S(\lmast) &=& \beta_1 S(\lmasd \lmasu \lmasd) + \beta_2 \{ S(\lmasd \lmasu) +
\\
           & & S(\lmasu \lmasd) + S(\lmasd \lmasd) \} + \beta_3 S(\lmasd) +
\beta_4 S(\lmeu)
            \nonumber \\
S(\lmet) &=& \beta_1 S(\lmasd \lmasu \lmasd ) + \beta_2 \{ S(\lmasu \lmasd   )
         \nonumber  \\
           & & + S(\lmasd \lmasu ) + S(\lmasu \lmasu) \} + \beta_3 S(\lmasu) +
\beta_4 S(\lmed)
           \nonumber \\
S(\lzt) &=& \beta_5 \left\{ S(\lmast) + S(\lmet) \right\} +
\beta_6 S(\lzu \lzd  )
\nonumber
\end{eqnarray}
\end{center}

%S(\ltimes) &=& \alpha_1 S(L_{+}) + \alpha_2 S(L_{-}) + \alpha_3 S(L_{0})
%\nonumber \\
%S(b^{+}) &=& \beta_1 S(g_{2} g_{1} g_{2}) + \beta_2 \{ S(g_2 g_1 ) +
%\nonumber \\
%           & & S(g_1 g_2) + S(g_{2}^{2}) \} + \beta_3 S(g_2) + \beta_4
%S(g_{1}^{-1})
%            \nonumber \\
%S(b^{-}) &=& \beta_1 S(g_2 g_1 g_2) + \beta_2 \{ S(g_2 g_1) \label{eq3}
%         \label{next} \\
%           & & + S(g_1 g_2 ) + S(g_{1}^{2}) \} + \beta_3 S(g_1) + \beta_4
%S(g_{2}^{-1}) \nonumber \\
%S(b^{0}) &=& \beta_5 \left\{ S(b^{+}) + S(b^{-}) \right\} + \beta_6 S(I_{1}
%I_{2})\nonumber

%S(A_i) &=& \alpha_1 S(G_i) + \alpha_2 S(G_i^{-1}) + \alpha_3 S(I_i)  \\
%S(B_i^{+}) &=& \beta_1 S(\Gi\Gime\Gi) + \beta_2 \{ S(\Gi\Gime)   \nonumber \\
%           & & + S(\Gime\Gi) + S(\Gi^{2})\} + \beta_3 S(\Gi) + \beta_4
%S(\Giime) \nonumber \\
%S(B_i^{-}) &=& \beta_1 S(\Gi\Gime\Gi) + \beta_2 \{ S(\Gi\Gime)  \nonumber \\
%           & & + S(\Gime\Gi) + S(\Gime^{2})\} + \beta_3 S(\Gime) + \beta_4
%S(\Gii) \nonumber \\
%S(\Bzei) &=& \beta_5 \left\{ S(\Bmasi) + S(\Bmei) \right\} + \beta_6
%S(I_{i-1} I_{i}) \nonumber

Obviously an ambient invariant HOMFLY polynomial for  intersecting
loops may be defined by means of eq. \ref{next}. It should be noted that our
 generalized $S_{L}$ polynomial contains as a particular case the generalized
Jones Polynomial
of \cite{Pullin} if there are no 6-valent vertices $\alpha_1 =
q^{1/4}(1-\epsilon)$,
$\alpha_2 = 0$, $\alpha_3 = \epsilon$, $z=q^{1/2}-q^{-1/2}$, $a = q^{3/4}$ and
$b = q^{1/4}.$
Now  we shall connect this polynomial with the
Wilson loop
average of
the Chern Simon's Theory for intersecting loops.

% \section{\bf Quantum Chern Simons Theory and intersecting link invariants.}

It is well known \cite{witten} \cite{guad} \cite{smolin} that
 the Wilson loop  expectation  value  of  the
Chern Simons (CS) theory are link invariants. In particular,  if  the
gauge group is SU(N) they  correspond  to  the  regular  invariant
HOMFLY polynomial with certain choice of their variables. Our  aim
in what follows is to show that this relation holds true even when
the links  have 4 or  6-valent   intersections.  Our  approach  is
variational and it was worked out  by  Smolin  \cite{smolin},
 Guadagnini et al. \cite{guad} and
extended for 4-valent  vertices  by  B.  Brugmann,  R.Gambini  and
J.Pullin \cite{gamPu}.
We will consider the CS theory with gauge  groups  SU(N)  in
the  fundamental representation. The Wilson loop expectation value
is given by

%EQUATION 1
\begin{equation}
\langle W(\gamma) \rangle = Z^{-1} \int  W(\gamma) e^{iS_{CS}[A]} \  {\cal D}A
\end{equation}

%EQUATION 2
\begin{equation}
S_{CS}[A] = k/4\pi \int \epsilon^{\mu\nu\rho} {\cal T}r[A_{\mu} \partial_{\nu}
A_{\rho} + i\frac{2}{3}A_{\mu}A_{\nu}A_{\rho}] \ d^3 x
\end{equation}

%EQUATION 3
\begin{eqnarray}
W(\gamma) = {\cal T}r[P \ e^{i\oint_{\gamma} A_{\mu} \ dx^{\mu} }] \\
U(\gamma_{x}^{y}) =  P \ e^{ i{ {\int_x^y}_{{ }_\gamma} }  A_{\mu}  \ dx^{\mu}
 }
\end{eqnarray}

Two main relations are necessary to derive the skein relations for
$\wilson$ , the first one allows to write the field strength in terms of
the functional derivative of the CS action.

%EQUATION 4
\begin{equation}
F^{a}_{\mu\nu}(x) e^{iS_{CS}} = \frac{-i4\pi}{k} \epsilon_{\mu\nu\rho}
   \frac{\delta}{\delta A_{\rho}^{a}(x)} e^{iS_{CS}}
\end{equation}
and the second, to express a deformation of the Wilson loops

%EQUATION 5
\begin{equation}
\delta W(\gamma) = i d x^{\mu} dy^{\nu}F_{\mu\nu}^{a}(x){\cal T}
r[T^aU(\gamma_{x}^{x})]
\end{equation}
where the $T^{a}$'s are the generators of $ SU(N)$, $\dx$ ,$\dy$  expand
the area element  of the deformation, and the generators have  been
inserted at the point  x  of  the  loops.  We  will  perform   the
calculation in three steps, the first two parts  are  ensencially a
review of references \cite{smolin} \cite{guad}

If we add a small loop of area $ \dx \dy $ at the point  x of the
loop $\gamma$ we get using eq. 15.

%EQUATION 6
\begin{equation}
\delta W(\gamma) = Z^{-1} \int i d x^{\mu} dy^{\nu}F^{a}_{\mu\nu}(x){\cal T}
r[T^aU(\gamma_{x}^{x})] e^{iS_{CS}}  {\cal D}A
\end{equation}

Using eq. 14 and integrating by parts

%Equation 7
\begin{equation}
\delta \langle W(\gamma) \rangle =  Z^{-1} \frac{-i4\pi}{k}
  \int \delta(x-z)dx^{\mu}dy^{\nu}dz^{\rho}\epsilon_{\mu\nu\rho}
   {\cal T}r[T^aT^aU(\gamma_{x}^{x})] e^{iS_{CS}[A]} {\cal D}A
\end{equation}

where we have used that

%In text
\begin{equation}
 \frac{\delta U(\gamma_{x}^{x}) }{\delta A_{\rho}^{a}(x)} =
idz^{\rho}T^{a}U(\gamma_{x}^{x})\delta(x-z)
 \end{equation}

and $\dz$ is oriented along the tangent to $\gamma$. The signature
of the volume factor depends on the relative orientation of the
area element $\dx \dy$ and the infinitesimal tangent $\dz$

Following  the  Kauffman  convention \cite{Kauffman} we  regularize $k$ in
order to ensure
  that the volume  factor takes the values $\pm  1/2$
or $0$.
Then using that for SU(N)

\begin{equation}
T^{a}T^{a} =  \frac{N^2-1}{2N}I
\end{equation}

we have
\begin{equation}
\delta \wilson = \pm  \frac{N^2-1}{2N} (-\frac{2\pi i}{k}) \wilson
\end{equation}

or

\begin{equation}
\delta \wilson = 0
\end{equation}

These equations lead to the skein relations

\begin{equation}
\langle W(\hat L_{\pm}) \rangle  -
\langle W(\hat L_{0}) \rangle = \pm  \frac{N^2-1}{2N} (-\frac{2\pi i}{k})
\langle W(\hat L_{0}) \rangle
\end{equation}

or

\begin{equation}
\langle W(\hat L_{\pm}) \rangle  =
[1  \pm  \frac{N^2-1}{2N} (-\frac{2\pi i}{k}) ]
\langle W(\hat L_{0}) \rangle
\end{equation}

When the deformation is introduced  in  the  plane  containing  the
tangent vector it does not affect  the topology of  the
loops as it is shown by equation 21.

% ii)  DEFORMATIONS in a 4-valent vertices

We add now an infinitesimal closed loop at a 4-valent vertex in
the plane of one of the tangents (see fig. 3)

\begin{equation}
\wilson = Z^{-1} \int  {\cal T}r[U_{23}(\gamma_{x}^{x})U_{41}(\gamma_{x}^{x})
]\chern \da
\end{equation}

and we obtain

\begin{eqnarray}
&&\delta \wilson =  \\
&& Z^{-1} \int  (\frac{-4\pi}{k})  \ep \dx \dy \deriv
{\cal T}r[T^a U_{23}(\gamma_{x}^{x}) U_{41}(\gamma_{x}^{x})]
\chern \da  \nonumber
\end{eqnarray}

where we have used eq. 14 and integrated by parts.  The action of the
functional derivatives vanishes in the segment 1-2, since the area
element $\dx \dy $ is in the plane of the tangent, and the result
is

\begin{eqnarray}
&& \delta \wilson  =   \\  && Z^{-1} \int  (\frac{-4i\pi}{k})  \ep \dx \dy \dz
 \delta(x-z)  \nonumber \\
&& {\cal T}r[T^a U_{23}(\gamma_{x}^{x})T^{a}U_{41}(\gamma_{x}^{x})]
\  \  \  \ \chern \da  \nonumber
\end{eqnarray}

Using the Fierz identity for the generators of $SU(N)$

\begin{equation}
T_{ij}^{a}T_{kl}^{a} = \frac{1}{2} \delta_{il} \delta_{jk} - \frac{1}{2N}
\delta_{ij} \delta_{kl}
\end{equation}

we get

\begin{eqnarray}
&& \delta \wilson = \nonumber \\
&& \frac{1}{2} \{ Z^{-1} \int  (\frac{-4i\pi}{k})  \ep \dx \dy \dz
 \delta(x-z)  \nonumber \\
 && {\cal T}r[U_{23}]{\cal T}r[U_{41}] \chern \da \} - \\
 && \frac{1}{2N} \{ Z^{-1} \int  (\frac{-4i\pi}{k})  \ep \dx \dy \dz
 \delta(x-z)   \nonumber \\
 &&  {\cal T}r[U_{23}U_{41}]] \chern \da \} \nonumber
 \end{eqnarray}

we can interpret this as

\begin{eqnarray}
\wilsonpm - \wilsont  = \pm \frac{1}{2}(-i2\pi/k) \wilsonz
\pm \frac{1}{2N}(-i2\pi/k) \wilsont
\end{eqnarray}

To first order in $1/k$ the equation 29 is equivalent to

\begin{eqnarray}
\wilsont  =  [1 \pm  \frac{1}{2N}(-i2\pi/k)] \wilsonpm
\pm \frac{1}{2}(-i2\pi/k) \wilsonz
\end{eqnarray}

and hence

\begin{eqnarray}
[1+ \frac{1}{2N}(-i2\pi/k)]\wilsonmas - [1- \frac{1}{2N}(-i2\pi/k)]
\wilsonmenos = \nonumber \\
(-i2\pi/k)\wilsonz
\end{eqnarray}

% iii) DEFORMATION ON A 6 VALENT VERTEX

We are ready to discuss the case of 6-valent vertices. As well as
in the other cases  we will consider  a deformation that breaks the
intersection.
We will restrict ourselves to one of the possible deformation of a
$b^{+}$ type intersection. The other cases, .i.e. deformations  in
$b^{-} $ or $b^{0}$, involve similar computations.
We consider the Wilson loop (see fig. 4)

\begin{eqnarray}
\wilson = \integral  {\cal T}r[U_{25}U_{63}U_{41}] \chern \da
\end{eqnarray}

Hence the variation under the small deformation is

\begin{eqnarray}
&& \delta \wilson = \nonumber \\
&& \integral (4\pi/k) \dx \dy \ep \deriv {\cal T}r[T^{a}U_{25}
U_{63}U_{41}] \chern \da = \\
&& \integral \cuatro \dx \dy \dz \ep  \delta(x-z) {\cal T}r[T^{a}U_{25}
U_{63}T^{a}U_{41}] \chern \da \nonumber
\end{eqnarray}

where we use eq. 14 to integrate by parts and note that the only one
term resulting from the functional derivative  doesn't vanish. Using
once again the Fierz identity and with the same convention for the
volume factor we get

\begin{eqnarray}
&& \delta \wilson = \nonumber \\
&& \mp  \frac{1}{2N} \left( \dos \right) \wilson \pm \\
&& \frac{1}{2}  \left( \dos \right)
\{ \integral \tra [U_{25}U_{63}] \tra [U_{41}] \chern \da \} \nonumber
\end{eqnarray}

We have verified that if one splits the intersection in the second
term of equation 34 by eliminating the kinks, its variation
vanishes. This allows us to interpret  equation 34 as

\begin{eqnarray}
\wilsonr - \wilsonb = \mp \frac{1}{2N} \left( \dos \right) \wilsonb \pm
\frac{1}{2} \left( \dos \right) \wdume
\end{eqnarray}

where

\begin{eqnarray}
R_{+} &=& \ltimesd \ltimesu \lmasd      \nonumber \\
R_{-} &=& \lmeu \ltimesd \ltimesu
\end{eqnarray}

Finally from equation 35 we obtain to first order in $1/k$

\begin{eqnarray}
 \wilsonb = [1 \pm  \frac{1}{2N} (\dos)] \wilsonr \mp
\frac{1}{2} (\dos) \wdume
\end{eqnarray}

At this point we make contact with the results of the first part of
this letter. As it
has been already pointed out equations 23 and 31 are consistent to first
order in $1/k$ with  the skein relations

\begin{eqnarray}
\wilsonpmh &=& q^{\frac{N^{2}-1}{2N}} \wilsonzh \\
q^{\frac{1}{2N}} \wilsonmas - q^{\frac{-1}{2N}} \wilsonmenos &=&
(q^{1/2}-q^{-1/2}) \wilsonz \nonumber
\end{eqnarray}

where  $ q= e^{\dos}.$ This means that a suitable normalization
of $\wilson$  corresponds to  a specialization of $S(L)$ polynomial in the non
intersecting case. We have

\begin{equation}
\frac{\langle W(L) \rangle}{\langle W(U) \rangle} =  S_{L}(
a= q^{\frac{N^{2}-1}{2N}},b= q^{\frac{1}{2N}},z=q^{\frac{1}{2}}
-q^{ \frac{-1}{2}})
\end{equation}

On the other side equation 30 corresponds to the equivalent skein
relations

\begin{eqnarray}
\wilsont = q^{1/2N} \wilsonmas + (1-q^{1/2}) \wilsonz \\
\wilsont = q^{-1/2N} \wilsonmenos + (1-q^{-1/2}) \wilsonz
\end{eqnarray}

Also we get from equation 30 and equation 38

\begin{eqnarray}
\wilsont = \frac{1}{q^{1/4}+q^{-1/4}}[ q^{-\frac{N-2}{4N}} \wilsonmas +
 q^{\frac{N-2}{4N}} \wilsonmenos]
\end{eqnarray}

Hence

\begin{eqnarray}
\alpha_1 = \frac{ q^{-\frac{N-2}{4N}} } {q^{1/4}+q^{-1/4}} \\
\alpha_2 = \frac{ q^{\frac{N-2}{4N}} } {q^{1/4}+q^{-1/4}}
\end{eqnarray}

In the case of 6-valent vertices we have eq. 37

\begin{eqnarray}
\wilsonb = \delta_1^{\pm}  \wilsonr \pm \delta_2^{\pm} \wdume
\end{eqnarray}

with

\begin{eqnarray}
\delta_1^{\pm} = 1 \pm \frac{1}{2N} \left( \dos \right) \\
\delta_2^{\pm} =  \mp \frac{1}{2} \left( \dos \right)
\end{eqnarray}

Besides form equations 30 and 31   we get

\begin{eqnarray}
\wilsongii = \gamma_1 \wilsongi + \gamma_2 \wilsoni \\
\wilsona = \phi_1 \wilsongi + \phi_2 \wilsoni
\end{eqnarray}

where

\begin{eqnarray}
\gamma_1 =  1 +  \frac{1}{N}( \dos)  \\
\gamma_2 = - \left( \dos \right) \\
\phi_1 =  1 +  \frac{1}{2N} (\dos)  \\
\phi_2 = - \frac{1}{2} (\dos)
\end{eqnarray}

And from equations  36, 45, 48 and 49 it results

\begin{eqnarray}
\wilsonb &=& \delta_1^{+} \phi_1^{2} \wilsongfg
 + \delta_1^{+}\phi_1 \phi_2 [\wilsongg + \wilsonx] + \nonumber \\
& &  \delta_2^{+} \gamma_1 \wilsony
 + [\delta_1^{+}\phi_2^{2} +
\delta_2^{+}\gamma_2 ] \wilsongi
\end{eqnarray}

and

\begin{eqnarray}
\wilsonb &=& \delta_1^{-} \phi_1^{2}\gamma_1 \wilsongfg
+ [\delta_{1}^{-}\phi_{1}^{2}\gamma_{2}+\delta_{2}^{-}\gamma_1] \wilsony
\nonumber \\
& &  + \delta_1^{-}\phi_1 \phi_2 \gamma_1 \wilsonx +   \\
& & [\delta_1^{-}( 2 \alpha_1 \alpha_2 + \alpha_2{2}) + \delta_{2}^{-}
]\gamma_2
\wilsongi
+  [\delta_1^{-} (\phi_1 \phi_2 + \phi_2^{2}) \gamma_1] \wilsongg \nonumber
\end{eqnarray}

By consistency between equations 54 and 55 and with the fifth equation of eq.
9
 it
must be to first order in $1/k$

\begin{eqnarray}
\beta_1 &=& \delta_1^{+} \phi_1^{2} =  \delta_1^{-} \phi_1^{2}\gamma_1 \\
\beta_2 &=& \delta_1^{+}\phi_1 \phi_2 =  \delta_{2}^{+} \gamma_1 =
\delta_1^{-} \phi_1^{2} \gamma_2 + \delta_2^{-}\gamma_{1} =\nonumber \\
 && \delta_1^{-} \phi_1 \phi_2 \gamma_1 = \delta_1^{-} ( \phi_1 \phi_2 +
\phi_2^{2}) \gamma1 \\
\beta_3 &=& [\delta_1^{+}\phi_2^{2} +
\delta_2^{+}\gamma_2 ] = [\delta_1^{-} (2\phi_1 \phi_2 + \phi_2^{2}) +
\delta_2^{-}]\gamma_2
\end{eqnarray}

This equations are actually satisfied and we get

\begin{eqnarray}
\beta_1 = 1 + \frac{3}{2N}(\dos) + {\cal O}(1/k^{2})  \nonumber \\
\beta_2 =  -\frac{1}{2}(\dos) + {\cal O}(1/k^{2}) \\
\beta_3 = {\cal O}(1/k^{2}) \nonumber
\end{eqnarray}

Since our computation was pertubative in $1/k$ we  cannot  compute
the exact form of the $\beta$'s,  but the  equations 56, 57, and 58, strongly
suggest  that  the  relation  between  $ \wilson$  and  the  regular
invariant polynomial $S_{L} $ also holds in the intersecting case.
Now we would like to make contact  with  the
relation found by B.Brugmann, R. Gambini and J.Pullin \cite{Pullin} in quantum
gravity with a cosmological constant.
They proved that the regular invariant polynomial $S_{L}$
associated to the CS theory was a solution of the full set of
constraints of quantum gravity with cosmological constant
$\Lambda$ in the loop representation provided that $ k= i24\pi/\Lambda$.
 Their computation was restricted to the case of 4-valent
intersections. Our work allows to extend this result to 6-valent
vertices. As it was already mentioned in the first part, they are
required to show the  existence of non degenerate quantum metrics.
As it was recently shown by J. Baez \cite{Baez2} in the case of tangles,
link polynomials
may be used as the starting point for the definition of an inner product in
Quantum Gravity. Our results could allow to define the inner product of states
which depend  on intersecting loops. On the other hand, the Chern Simon
perturbation
theory leads to certain analytical expression for the knot invariants
coefficients
(Gauss number, second Alexander-Conway coefficient, etc.). These expressions
are framing dependent in the intersecting case. The correspondence between
these coefficients and the HOMFLY polynomials allow to introduce a consistent
framing of this quantities  that play a fundamental role in Quantum Gravity.
\cite{Pullin}

% ************************************* 31
%\subsection{\bf The solution of the differential constraint}

% ************************************* 4
%\section{\bf Free coordinates on the loop space}
% ************************************* 5
%\section{\bf Knot invariants}

% ************************************* 6
%\section{\bf Conclusions}

\section*{\bf Acknowledgments}

We wish to thank Ricardo Siri for many fruitful discussions.
%R. Gambini wishes to thank Abhay Ashtekar, Lee Smolin, Jorge Pullin and
%Bernd Br\"ugmann  for fruitful discussions.

% ************************************* REFERENCES

\newpage
\section*{Figure Captions}
\begin{description}
\item [fig. 1] Graphical representation of the braid generators.
\item [fig. 2] Graphical representation of new crossings in the planar knot
diagrams.
\item [fig. 3] Deformation that breaks a 4-valent vertex.
\item [fig. 4] Deformation that breaks a 6-valent $b^{+}$ type vertex.
\end{description}

\end{document}